# Enhancing the Performance of CdSe/CdS Dot-in-Rod Light Emitting Diodes via Surface Ligand Modification

*Prachi Rastogi[a,b], Francisco Palazon[a], Mirko Prato[c], Francesco Di Stasio[a,*] and Roman Krahne[a,*]*

[a]Nanochemistry Department, Istituto Italiano di Tecnologia, Via Morego 30, 16163 Genova, Italy

[b]Dipartimento di Chimica e Chimica Industriale, Università degli Studi di Genova, Via Dodecaneso 31, 16146 Genova, Italy

[c]Materials Characterization Facility, Istituto Italiano di Tecnologia, Via Morego 30, 16163 Genova, Italy

* corresponding author email: roman.krahne@iit.it, francesco.distasio@iit.it

ABSTRACT

The surface ligands on colloidal nanocrystals (NCs) play an important role in the performance of NCs based optoelectronic devices such as photovoltaic cells, photodetectors and light emitting diodes (LEDs). On one hand, the NC emission depends critically on the passivation of the surface to minimize trap states that can provide non-radiative recombination channels. On the other hand, the electrical properties of NC films are dominated by the ligands that constitute the barriers for charge transport from one NC to its neighbor. Therefore, surface modifications *via* ligand-exchange have been employed to improve the conductance of NC films. However, in light emitting devices, such surface modifications are more critical due to their possible detrimental effects on the emission properties. In this work, we study the role of surface ligand modifications on the optical and electrical properties of CdSe/CdS dot-in-rods (DiRs) in films, and investigate their performance in all-solution processed LEDs. The DiR films maintain high PLQY, around 40-50 %, and their electroluminescence in the LED preserves the excellent color purity of the PL. In the LEDs, the ligand-exchange boosted the luminance, reaching a four-fold increase from 2200 cd/m$^2$ for native surfactants to 8500 cd/m$^2$ for the exchanged aminoethanethiol ligands. Moreover, the efficiency roll-off, operational stability, and shelf life are significantly improved, and the external quantum efficiency is modestly increased from 5.1 % to 5.4 %. We relate these improvements to the increased conductivity of the emissive layer, and to the better charge balance of the electrically injected carriers. In this respect, we performed ultraviolet photoelectron spectroscopy (UPS) to obtain deeper insight in the band alignment of the LED structure. The UPS data confirms similar flat-band offsets of the emitting layer to the electron- and hole-transport layers, respectively, in the case of aminoethanethiol ligands, which translates to more symmetric barriers for charge injection of electrons and holes. Furthermore, the change in solubility of the nanocrystals induced by the ligand-exchange allows for a layer-by-layer deposition process of the DiR films that yields excellent homogeneity and good thickness control, and enables the fabrication of all the LED layers (except for cathode and anode) by spin-coating.

**Introduction**

Colloidal semiconductor nanocrystals (NCs) are an unique optical material with many fascinating characteristics such as wide emission wavelength tunability,[1] narrow bandwidth emission[2] and high photoluminescence quantum yield (PLQY),[3] which makes them a promising material for light emitting applications.[4-7] Furthermore, NCs can be fabricated and processed via low-cost solution based techniques that are compatible with lightweight and flexible substrates.[8,9] Owing to these unique features, NCs have been widely investigated for their use in optoelectronic devices, for example, light-emitting diodes (LEDs),[5,10-12] lasers,[13-15] solar concentrators,[16,17] photodetectors[18,19] and photovoltaics.[20] Among these potential applications, the realization of high performance NC LEDs has been of particular interest due to their potentially significant impact in display and lighting technologies. After the first demonstration of NC LEDs,[21] there have been numerous attempts to improve the device efficiency and luminance by optimizing both materials[7,22-24] and device architecture.[10,25-27]

In general, the external quantum efficiency (EQE) of an LED depends on the charge carrier mobility and injection rate, on the optical out-coupling, and on the PLQY of the active NC layer. The presence of ligands on the surface of NCs plays an important role in determining the electronic properties[28] of the NCs and affects the performance of the device.[29,30] The ligands passivate the surface of the NCs, which directly impacts the density and depth of trap states, thus affecting charge transport. Typically, the surface ligands represent tunnel barriers for the charge transfer from one NC to the next, which determines the charge carrier mobility of the film. Concerning nanocrystal synthesis, the surface ligands provide essential colloidal stability and solution processability. This task is well fulfilled by bulky ligands like trioctylphosphine oxide (TOPO) or octadecylphosphonic acid (ODPA), which, however, are detrimental for optoelectronic applications. Therefore, strategies to exchange those long-chain ligands with shorter ones have been developed.[31,32] Such ligand-exchange treatment on colloidal NCs has been successfully employed in solar cells to improve the device performance.[33-35] In devices that rely on the forward injection of electrons and holes, such as LEDs, high mobility contributes to efficient exciton formation. However, applying such ligand-exchange processes to NC based LEDs is challenging, since it degrades typically the PLQY,[36] which affects also the internal quantum efficiency of the device. Therefore, improved strategies for the ligand-exchange, as well as a good understanding of the influence of the surface ligands on the LED performance is crucial for the development of NC LEDs.

In this work, we investigate the effect of different surface ligands on the performance of LEDs based on CdSe/CdS "dot-in-rods" (DiRs) as emissive layer. We chose three different molecules to replace the native ODPA surface ligands, namely mercaptopropionic acid (MPA), aminoethanethiol (AET), and ammonium thiocyanate (SCN). The structure of these molecules is shown in Figure S1 of the Supporting Information. They differ in size, which should affect the conductivity of the emissive layer, and in charge, which should influence the band alignment. CdSe/CdS DiRs were selected as emitters, because they are an attractive material for LEDs, since they manifest high PLQY, the emission wavelength can be tuned via the core size over a large portion of the visible range, and the rod shape can induce polarized emission.[15] The ligand-exchange was applied to DiR films in a layer by layer (LbL) deposition process that yielded better results in terms of PLQY and layer quality than ligand-exchange performed in solution (see Figure S2). Characterization of the optical and electrical properties of the DiR films confirmed PLQY values in the range of 38-48 %, and demonstrated increased conductivity due to the ligand-exchange with smaller molecules. Studying the differently functionalized DiR films in LEDs revealed that the LED efficiency, luminance, stability and shelf lifetime, as well as the efficiency roll-off were strongly influenced by the type of surface ligands. For the best performing ligand, we achieved a small improvement in external quantum efficiency (EQE), but more importantly, we were able to dramatically improve the luminance (reaching 8500 cd/m$^2$ at 638 nm), efficiency roll-off and stability of the device.

**Results & Discussion**

The synthesis of octadecylphosphonic acid (ODPA) capped CdSe/CdS DiRs (TEM image in Figure 1a) was carried out according to the seeded growth approach at high temperature reported by Carbone et al.,[37] with slight modifications. In particular, we increased the growth temperature from 365°C to 380°C and the CdO:S amount from 60:60 mg to 90:90 mg, which led to DiRs with a thick CdS shell. Figure 1a shows a TEM image of the obtained rods that had a diameter of 9 nm and a CdSe core with 4.3 nm diameter (see Methods section for synthesis details). The relatively thick CdS shell is favorable to maintain high PL intensity and PLQY also in DiR films on which ligand-exchange processes were applied.[36] The optical properties of the ODPA capped DiRs in solution are reported in Figure S2 of the SI, manifesting a single, Gaussian PL peak centered at 632 nm and a PLQY of 60%. For our targeted application in LEDs, the DiRs as light emitting material have to be implemented as a thin film. Therefore, the optical properties of the DiRs in such thin films are of importance, and not necessarily their performance in solution. To this end, we have fabricated DiR thin films by spin coating of the DiRs from toluene suspensions, with particle concentration of 22 mg/mL that was necessary to obtain homogenous films. We have performed the ligand-exchange process on such films in order to investigate the effect of different surface ligands on the optical properties of the DiR films (and eventually on the performance of DiR based LEDs). In this respect, the ligand-exchange on DiR films, as opposed to the DiRs in solution, has the advantage that it enables a LbL deposition procedure, which allows to fabricate homogenous, and void-free films with excellent thickness control. Such LbL procedure is well established for PbS NC solar cells,[38] but so far in-situ ligand-exchange procedures have only been marginally explored in the fabrication of nanocrystal based LEDs.[39,40] In detail, a layer of ODPA capped DiRs is spincoated on the substrate, and then the ligand-exchange is applied to the film from methanol or acetone solutions. This not only changes the surface chemistry of the NCs, but also renders the film insoluble in toluene, and therefore allows for subsequent steps of spin coating DiRs from toluene, as is illustrated in more detail in Figure S3 in the SI. Typically, we deposited eight layers *via* this LbL process. For our study, the pristine ODPA ligands are exchanged in films with MPA, AET or SCN. The roughness of films obtained with the LbL process was evaluated by atomic force microscopy (see Figure S3). Fourier-Transform-infrared (FTIR) spectroscopy data from films before and after ligand exchange are reported in Figure S4. Figure 1b shows a scanning electron microscopy (SEM) image of a DiR film where AET ligand-exchange was applied. The steady state photoluminescence (PL) spectra of films with ODPA capped and ligand exchanged DiRs are shown in Figure 1c, and demonstrate that the procedure does not lead to drastic changes in the emission of the DiR film: the emission wavelength is maintained ($\lambda$ = 633 nm for all films), and QYs are in the range from 38 to 48 % as detailed in the legend of Figure 1. The PL decay recorded at the emission maximum is shown in Figure 1d. Fitting of the decay curves with a bi-exponential function allows to extract average decay times, and we obtain 11.8 ns for the ODPA capped DiR films, 14.1 ns for SCN, 17.7 ns for MPA, and 18.1 ns for AET. The increase in average lifetime caused by the ligand exchange coincides with a reduction in PLQY, which indicates that the specific properties of the surface ligands affect both the radiative and the non-radiative lifetime components. By combining the QY values with the average decay times (see equations (1) and (2) in the SI), we can calculate the average radiative decay rates ($\Gamma_{RAD}$) for the films with the different ligands. We find that the radiative decay rate is more strongly reduced in the ligand-exchanged films than the non-radiative decay rate (with respect to the ODPA ones), which points to less efficient surface passivation, as could be expected. Table 1 summarizes the optical properties of the different DiR films.

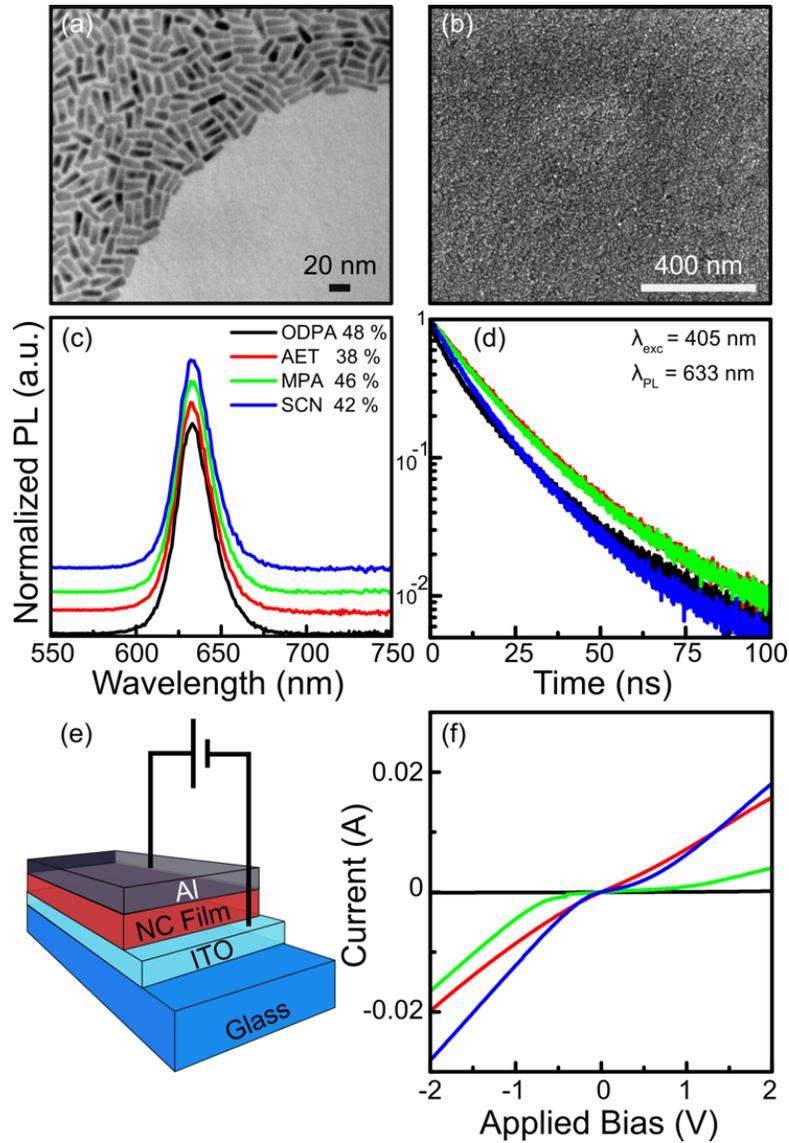

Figure 1 (a) Transmission electron microscope (TEM) image of CdSe/CdS Dot-in-Rods (average length = 17 nm, width = 9 nm and CdSe core diameter = 4.3 nm). (b) Scanning electron microscope (SEM) image of a AET ligand-exchanged DiR film. (c) Steady-state PL spectra of pristine ODPA capped DiRs and ligand-exchanged DiR films. (d) PL decay traces from pristine and ligand-exchanged DiRs films, with excitation wavelength at 405 nm. Color legend in (c) applies also to (d,f). (e) Sample architecture for measuring the DiR film conductivity. (f) Current-voltage curves for the different films.

Table 1. Average decay times, QY, the calculated radiative and non-radiative average decay rates, and the current and conductivity taken at a bias of -2V of the different films.

|  | $\tau_{AV}$ (ns) | QY % | $\Gamma_{RAD}$ ($\mu s^{-1}$) | $\Gamma_{NON-RAD}$ ($\mu s^{-1}$) | I (A) | $\sigma$ (S/cm) |
|---|---|---|---|---|---|---|
| ODPA | 11.8 | 48 | 40.7 | 44.1 | -1.20 $10^{-5}$ | 9.60 $10^{-10}$ |
| SCN | 14.1 | 42 | 29.7 | 41.1 | -0.027 | 2.16 $10^{-6}$ |
| MPA | 17.7 | 46 | 26.0 | 30.6 | -0.016 | 1.28 $10^{-6}$ |
| AET | 18.1 | 38 | 21.0 | 34.2 | -0.020 | 1.60 $10^{-6}$ |

The electrical conductivity of the DiR films depends strongly on the surface functionalization and therefore on the different ligands. We have measured the conductivity of the DiR films in vertical configuration with ITO and Al contacts, as shown in Figure 1 e,f. We find that the film conductivity increases with decreasing size of the ligand molecules, which can be related to the reduced inter particle spacing.[41] The conductivity values for similar measurement geometries are reported in Table 1, showing best performance for SCN, followed by AET and then MPA. ODPA capped DiR films show the lowest conductivity, as expected.

The architecture of the LED, in which the DiR films are integrated, is illustrated in Figure 2a. It employs ZnO nanoparticles (NPs) as the as electron-transport layer (ETL) and poly[N,N'-bis(4-butylphenyl)-N,N'-bisphenylbenzidine] (poly-TPD) in combination with polyvinyl-carbazole (PVK) as hole transport layer (HTL). After optimization of the thickness of the different layers, we obtained the following structure for the ODPA-based LED: patterned ITO / PEDOT:PSS (35 nm)/ Poly-TPD (35 nm) / PVK (7 nm) / DiRs film (28 nm) / ZnO NPs (100 nm) / Al (100 nm). We chose poly-TPD due to its high hole mobility ($\approx 10^{-4}$ cm$^2$V$^{-1}$s$^{-1}$),[42] and combined it with PVK because the deep highest occupied molecular orbital (HOMO) level at 5.8 eV of PVK[12] decreases the energy barrier for hole injection into the DiR layer. On the other side of the DiR layer, a thin film of colloidal ZnO NPs is employed as ETL because of the high electron mobility ($2\times 10^{-3}$ cm$^2$V$^{-1}$s$^{-1}$)[10] and its solution processability. Furthermore, the ZnO NP layer has an electron affinity of 4.1 eV, which is beneficial for the injection of electrons from the Al electrode into the active layer, while its high ionization potential of 7.4 eV and the resulting high VBM energy offset at the DiR-ZnO layer interface blocks the transport of holes.[10] Figure S6 shows a cross section SEM image of the LED structure, and sketches the band alignment of the layers.

We fabricated a LED structure based on the pristine, ODPA capped DiRs that serves as reference for comparison with those obtained with ligand-exchange DiR films. For the pristine, toluene soluble DiRs, the LbL process cannot be applied and the 28 nm thick DiR film is deposited in a single spin coating step. For this LED we obtain the following parameters: turn-on voltage of 2.2 V, luminance of 2200 cd/m$^2$, and a peak EQE of 5.1%. This device performance is comparable to the best reported values for DiR based LEDs with fully solution processed interlayers that report an EQE of 6.1%.[43]

Next, we replaced the DiR film in the above-mentioned architecture with those fabricated by the ligand-exchange LbL process. Figures 2d-f summarize the performance of the LEDs with the different surface ligands.

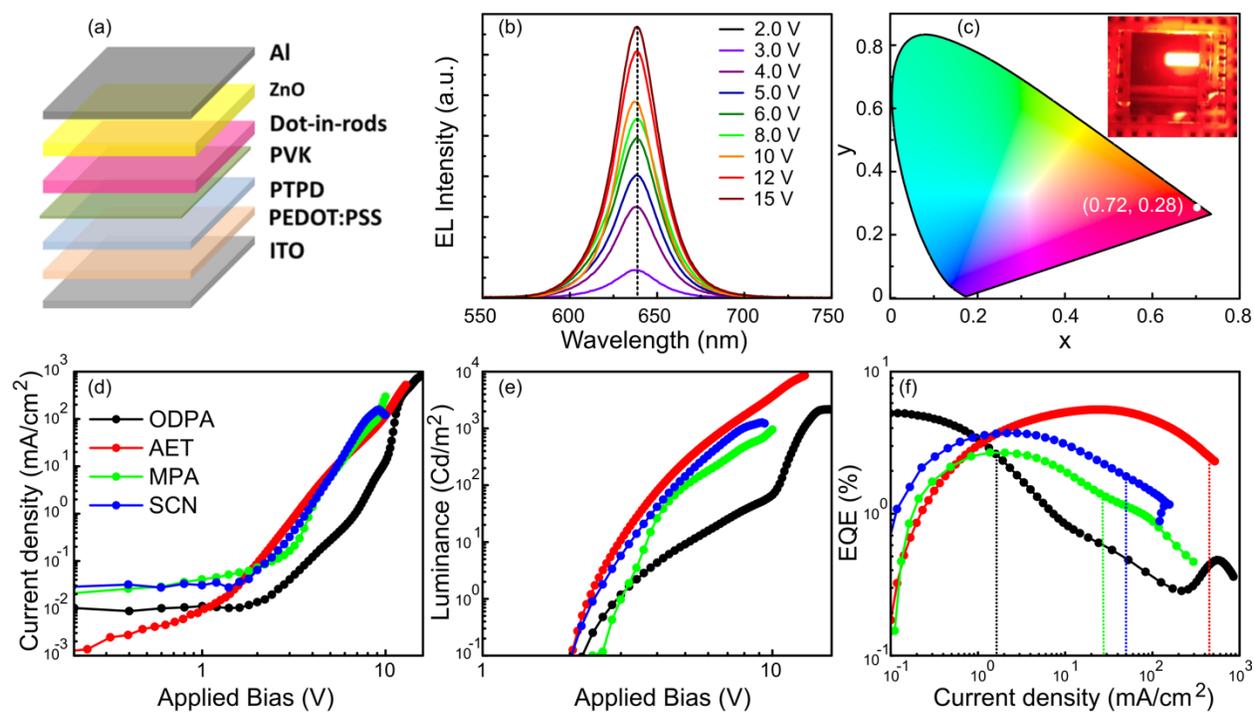

**Figure 2.** (a) Device structure of the multilayer LED. (b) EL spectra of a typical AET-DiR LED as a function of applied bias (c) CIE 1931 color coordinates representing the deep red emission of DiR LED (inset shows a photo of a LED at 10 V bias). (d-f) Characteristics of the LEDs with the DiRs passivated by different surface ligands. (d) Current density-voltage curves, (e) luminance-voltage characteristics, (f) and external quantum efficiencies (EQE). The dashed lines in (f) show the current at 50% of the peak EQE.

The electroluminescence (EL) spectra of a typical LED are plotted in Figure 2b for different bias voltages. We observe a single, Gaussian-shaped peak, centered at 638 nm. Importantly, neither the peak position nor its shape is affected by the increasing the bias voltage to high values. The color purity of the electroluminescence is demonstrated by the EL peak that manifests a narrow full width half maximum (FWHM) of 28 nm at a bias of 10 V (only slightly broader than the PL emission peak shown in Fig 1, see also Figure S2 for the PL from solution). At an emission wavelength of 638 nm, this corresponds to the color coordinates (x = 0.72, y = 0.28) of the Commission Internationale de l'Eclairage (CIE), thus presenting deep red color emission ideal for RGB display applications. We highlight that the brightness of the obtained peak (see Table 2) is particularly remarkable if one considers that the EL lies in the very low sensitivity tail of the human eye response.

Figures 2c-d report the current density-voltage and luminance-voltage (J-V-L) characteristics of the different LEDs. The ligand-exchange treatment with AET, MPA and SCN results in higher current densities, which can be directly related to the improved conductive properties of the respective ligand-exchanged DiR films shown in Figure 1f. However, only in the case of the AET ligands the performance of the LED in terms of EQE and peak luminance was improved, which in particular, led to a significant increase in luminance by a factor of 4, from 2200 cd/m$^2$ for ODPA to 8552 cd/m$^2$ for AET (see Figure 2e). Due to the better conductivity, the SCN and MPA coated DiR films show higher luminance at low bias voltage with respect to the pristine ODPA ligands. We note that adjustments of the DiR and ETL layers in the AET based LED to 40 nm and 85 nm, respectively, contributed to optimize ultimately their performance. The parameters of the different LEDs are summarized in Table 2.

Table 2. Parameters of the LEDs based on the DiR films with different ligands.

| Ligands | EQE % | J at EQE mA/cm$^2$ | J at EQE$^{1/2}$ mA/cm$^2$ | Luminance at EQE$^{1/2}$ Cd/m$^2$ | Peak luminance Cd/m$^2$ |
|---|---|---|---|---|---|
| ODPA | 5.1 | 0.1 | 1.7 | 30 | 2200 |
| AET | 5.4 | 24.5 | 375 | 7311 | 8552 |
| MPA | 2.7 | 2.0 | 28 | 96 | 953 |
| SCN | 3.8 | 2.15 | 41 | 570 | 1275 |

The EQE of the LEDs as a function of current density is shown in Figure 2f for the different ligands. Concerning the EQE, the efficiency roll-off is an important characteristic that needs to be avoided or minimized, since high efficiency at high luminance, and therefore at high current density, is desired. The reference LED based on ODPA ligands shows a pronounced efficiency roll-off, as the EQE decreases from the start with increasing current density, and reduces to 50% (referred as $EQE_{1/2}$) at a current density of $J_{EQE1/2}$=1.7 mA/cm². The LEDs based on ligand-exchanged DiR films demonstrate a significantly improved roll-off behavior, with a pronounced maximum of the EQE at a finite current density. For MPA and SCN, the $J_{EQE1/2}$ occurs at 28 mA/cm² and 72 mA/cm², respectively. The best performance is obtained for AET, where the roll-off is strongly reduced: a high EQE is sustained over a large current range with a maximum of 5.4 % at 25 mA/cm², resulting in a high $J_{EQE1/2}$=558 mA/cm². The values reported here are for the best typical device, see Figure S7 for statistics on the performance of ODPA and AET functionalized LEDs.

The result that the AET-exchanged LEDs are outperforming the other devices is somewhat surprising, since, based on PLQY and conductivity, SCN was the most promising candidate. However, while high PLQY and good conductivity of the material are essential requirements for optimal LED performance, also a balanced injection of electrons and holes into the emissive layer is crucial for high efficiency. This holds particularly for minimizing the EQE roll-off at high bias voltage or driving current. The best performance of the DiR films with AET ligands in terms of luminance and peak EQE indicates that their energy band structure is favorable for a balanced charge injection. Therefore, we performed ultraviolet photoelectron spectroscopy (UPS) on the different DiR films to get deeper insight into the band alignment. UPS allows to determine the Fermi energy level and valence band maximum (VBM) with respect to vacuum. In particular, the minimum kinetic energy of emitted photoelectrons (secondary electron cut-off; Figure 3a) corresponds to the Fermi level of the material, whereas their minimum binding energy (Figure 3b) indicates the difference between the VBM and the Fermi level (which is set at Binding Energy = 0). Combining this data with the optical band gap of the CdS shell, that can be extracted from the absorption spectrum in Figure SI2 to be 2.5 eV, allows to obtain the conduction band minimum (CBM). The relative positions of the VBM and CBM of the DiR film with the other layers in the LED structure allows us to rationalize the device performance in terms of charge injection balance.

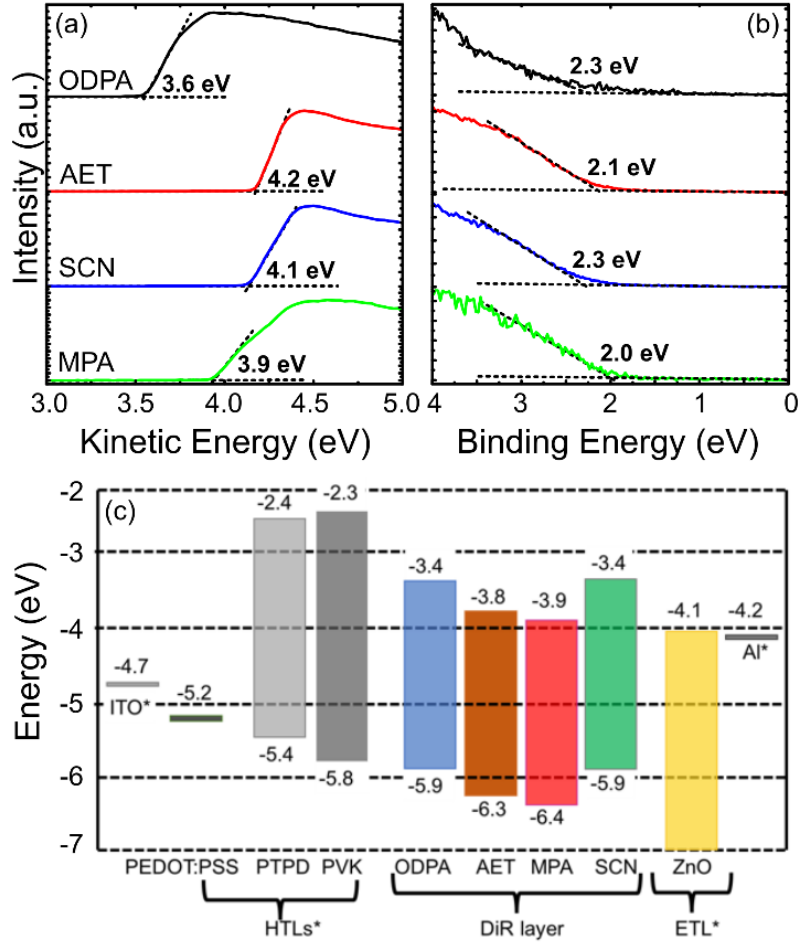

**Figure 3.** (a, b) UPS spectra obtained from the different DiR films. (c) Flat-band diagram of the LED structure. The energy bands of the differently functionalized DiR films are shown side-by-side for ease of comparison, however, only one layer is implemented in the LED at a time.

The UPS spectra of the films with the different ligands are shown in Figure 3a-b. We find that the ligand-exchange induced a significant shift of the vacuum to Fermi level energy difference (Figure 3a), from 3.6 eV for ODPA to 3.9 - 4.2 eV for the AET, MPA, and SCN films. For the VBM-to-Fermi energy (Figure 3b), we obtain values in the range from 2.0 to 2.3 eV; we have to note, however, that the assignment is less precise due to the soft onset of the low binding energy tail of the UPS spectrum.[15] Our results, combined with the CdS band gap of 2.5 eV, indicate that for all samples the Fermi level is much closer to the CBM than to the VBM. This behavior can be related to the peculiar band alignment of the CdSe and CdS within the DiRs, which manifests a large offset in the VB, and a small offset in the CB, *i.e.* the Fermi level position within the DiR band gap is closer to the CBM of the CdS shell.[44,45] While for the Fermi level the electronic structure of the complex DiR structure has to be considered, for the charge injection the band gap of the CdS shell is relevant, since the charge carriers have to pass the shell in order to recombine in the core region. The obtained values for the CBM and VBM for the DiR layers with the different ligands are depicted in the Figure 3c, together with the energy bands of HTLs, ETL, and the cathode and anode. The band scheme shows that the balance in barrier height for electrons and holes is best for AET (0.3 eV and 0.5 eV for electrons and holes, respectively), followed by MPA (0.6 eV; 0.2 eV), while in the case of ODPA and SCN electrons encounter a large barrier of 0.7 eV, and holes a much smaller barrier of 0.1 eV. Consequently, the AET-LEDs manifest the most equilibrated balance in charge carrier injection, which explains their best performance. A good charge balance in the emissive layer of the LED is of great importance for the following reasons.

On one hand, it directly affects the quantum efficiency, since it optimizes the fraction of carriers that can form excitons. Furthermore, it influences also the luminance, since the presence of excess charge carriers of one type will enhance the Auger recombination rate,[27] *i.e.* the non-radiative decay of excitons that are formed in DiR film. Furthermore, the accumulation of excess charge at the interfaces is detrimental to charge injection and further increases the charge imbalance. The latter effect is not pronounced at low current density, but increases with increasing bias and therefore strongly reduces the EQE and luminance at higher current density.

Summarizing this part, we find by comparing the band alignment with the LED performance, that the good conductivity of AET combined with a well equilibrated charge injection leads to the best performance in terms of EQE, luminance and efficiency roll-off. The similar behavior of the MPA and SCN passivated films in Figure 2d-e indicate compensation of the less equilibrated charge balance in SCN by the high conductivity, and vice versa for MPA. The better values of the SCN devices suggest that the film conductivity outweighs the charge imbalance. For ODPA, the low conductivity results in low current density at small bias voltage, which together with the high PLQY translates to good efficiency values, but poor luminance. Driving the ODPA LED at high bias leads to improvement in luminance, however is strongly at the expense of efficiency.

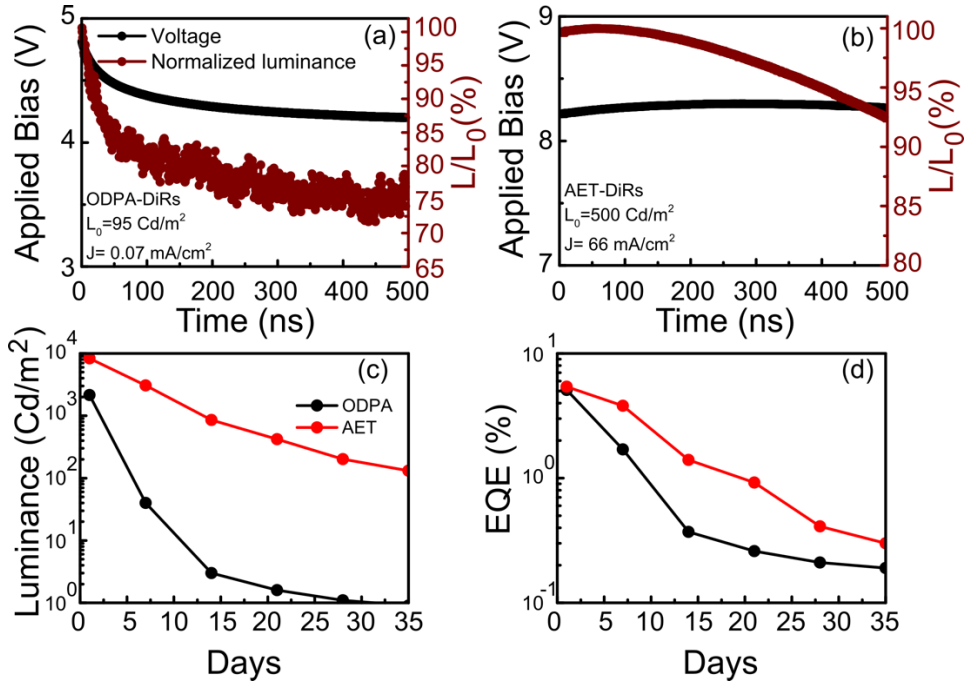

**Figure 4.** Luminance and driving voltage for (a) ODPA capped DiR LED and (b) AET capped DiR LED *versus* operating time under ambient conditions. Note the much higher driving current density and luminance (66 mA/cm$^2$ and 500 cd/m$^2$, respectively) in the case of AET with respect to ODPA (0.07 mA/cm$^2$ and 95 cd/m$^2$). (c-d) Luminance (c) and EQE (d) versus shelf time for ODPA and AET based devices.

Finally, we tested the stability of DiR LEDs, the operating lifetimes and shelf life under ambient condition at room temperature. All devices were encapsulated with epoxy glue and a cover glass inside a N$_2$-filled glove box before exposing them to air. The luminance and driving voltage versus time for the reference LED based on ODPA and the ligand-exchanged AET one, operated at a constant driving current density are shown in Figure 4a and b, respectively. For ODPA we obtain a luminance of 95 cd/m$^2$, while for AET we reach 500 cd/m$^2$. In addition to the much higher luminance at a similar bias voltage, the AET device also show a dramatically improved stability. While the luminance of the ODPA LED reduces to 75% of its initial value after 500 s, accompanied by a reduction in bias voltage, the AET LED maintains 93% of luminance over the same operation time, with only a marginal increase in bias voltage. Figure 4c and d depict the shelf life of the ODPA and AET LEDs by showing their luminance and EQE over days of storage time. Also in this case, a much slower degradation of the AET based LEDs is observed. The related luminance and EQE versus driving voltage curves are depicted in Figure S8.

**Conclusion**

In conclusion, the performance of LEDs based of films of colloidal NCs can be significantly improved by optimizing the surface chemistry *via* ligand-exchange. Such ligand-exchange can lead to much higher film conductivity, and impacts the energy levels of the NCs which can improve the balance of charge carrier injection in the emissive film, while the high PLQY of the films is preserved. In our study, based on CdSe/CdS DiRs, we obtained an enhancement in luminance from 2200 cd/$m^2$ for the pristine ODPA ligand to 8500 cd/$m^2$ for devices passivated with AET ligand, accompanied by a significant reduction in the efficiency roll-off, and improved stability and shelf time. The detailed characterization of the optical, electrical and energy level structure of the DiR films with different ligands allowed to identify the key properties and provided some insight on their influence on the device performance. This highlights that, apart from identifying the optimal inorganic semiconductor material as light emitter, its surface chemistry is of fundamental importance for the performance of the final LED.

**Materials and Methods**

**Synthesis of CdSe/CdS dot-in-rods**

CdSe QDs of diameter 4.3 nm were synthesized and purified according to the procedure described by Carbone et al.[37] For the shell growth, 90 mg of Cadmium oxide mixed together with 480 mg of ODPA, 3 gm TOPO, and 60 mg HPA were inserted in a flask that was then degassed at 130 °C for an hour. The shell growth was started by injection of 100 μM CdSe QDs dispersed in S:TOP (90 mg:1.5 gm) at 365 °C in the flask under nitrogen. After 8 minutes, the growth was stopped by cooling the reaction down. The samples were dispersed in a suitable solvent (toluene, chloroform), purified by precipitation with methanol, followed by centrifugation and re-suspension in toluene.

**Ligand-exchange in solution**

1 mL DiR solution with concentration of 2 mg/mL in toluene was added in 1 mL of 0.1 M mercaptopropionic acid (MPA): 0.12 M KOH-methanol solution. The solution was stirred for 10 minutes and then centrifuged at 4000 rpm for 3 minutes. Afterwards, the precipitate of ligand-exchanged DiRs was re-dispersed in water. For thiocyanate (SCN), the ligands were exchanged by adding 1 mL of 0.13 mM ammonium thiocyanate-methanol solution to 1 mL of DiR solution with concentration of 2 mg/mL in toluene. The mixture of ligands and DiRs was stirred for 5 minutes, followed by centrifugation and purification with isopropanol. The precipitate of SCN ligand-exchanged DiR was then redispersed in dimethyl sulfoxide (DMSO). The exchange with aminoethanethiol (AET) ligands was carried out with 0.05 M methanol solution of AET, which was then added to 1 mL of 2 mg/mL DiR toluene solution. The mixture of ligands and DiRs was stirred for 5 minutes. In the flocculated solution, 1 mL water was added to transfer AET ligand-exchanged DiRs in water. The water phase containing DiRs was separated, followed by centrifugation and purification with isopropanol. The AET precipitates were then re-dispersed in water.

**Ligand-exchange in films *via* layer-by-layer deposition**

The native ODPA ligands were replaced by MPA, AET and SCN via layer-by-layer (LbL) deposition. For the MPA ligand-exchange, the spin-coated DiR film was soaked in 1% (w/w) MPA in methanol for 30 s, and spin-dried at 2000 rpm for 45 sec. To remove the excess ligands, the film was washed with pure methanol on the spin coater at 2000 rpm (spin-washed). For the ligand-exchange to AET and SCN, 5 mg/mL AET methanol solution and 10 mg/mL SCN acetone solution were used, respectively. Afterwards, the AET and SCN exchanged films were spin-washed for two times, first with pure methanol and then with acetone. This procedure was repeated several times to achieve the desired thickness of DiR film. For instance, a 120 nm thick DiR film can be obtained with eight LbL depositions. The film thickness was evaluated by averaging several scans taken with a profilometer (Dektak).

**Steady state PL and time resolved PL**

Steady-state PL and time-resolved PL decay measurements were carried out on pristine and ligand-exchanged DiRs in solution and in films with an Edinburgh Instruments fluorescence spectrometer (FLS920), which included a Xenon lamp with monochromator for steady-state PL excitation and a time-correlated single-photon-counting unit coupled with a pulsed laser diode (excitation wavelength 405 nm) for time-resolved PL decay studies. The steady-state PL spectra recorded from films were obtained with an excitation wavelength of 450 nm. A calibrated integrating sphere was used to record the PLQY values from solution and from films. DiR solutions for PLQY measurements were prepared in quartz cuvettes and carefully diluted to 0.1 optical density at the excitation wavelength ($\lambda$ = 450 nm).

**Transmission electron microscopy**

Transmission electron microscopy (TEM) images were acquired on a JEOL JEM-1011 microscope equipped with a thermionic gun at 100 kV accelerating voltage. The samples were prepared by drop-casting diluted DiR colloidal suspensions onto 200-mesh carbon-coated copper grids.

**Ultraviolet-photoemission spectroscopy**

Ultraviolet photoemission spectroscopy (UPS) measurements have been carried out with a Kratos Axis Ultra$^{DLD}$ spectrometer using a He I (21.22 eV) discharge lamp. The analyses are conducted on an area of 55 μm in diameter, at pass energy of 10 eV and with a dwell time of 100 ms. The position of the Fermi level with respect to vacuum level is measured from the threshold energy for the emission of secondary electrons during He I excitation. A bias of −9.0 V is applied to the sample in order to precisely determine the low kinetic energy cut-off, as discussed in ref. [46]. Then, the position of the VBM versus vacuum level is estimated by the energy difference of the onset of the spectrum at low binding energy with respect Fermi level, according to the graphical method used in ref. [47]. The NC films for the UPS measurement have been deposited on Au-coated (thickness: 50 nm) 10 mm × 10 mm Si wafers.

**LED device fabrication and characterization**

LEDs were fabricated on patterned ITO glass substrates, which were cleaned in an ultrasonic bath using acetone and isopropanol sequentially. Prior to the PEDOT:PSS deposition, the ITO/glass substrates were further cleaned with oxygen plasma treatment for 10 mins at 15 W. The PEDOT:PSS layer was spin-cast on the cleaned ITO glass substrates at 4000 rpm, and dried at 140 °C for 10 minutes. As hole transport layer, poly TPD and PVK were dissolved in chlorobenzene and m-xylene at 8 mg/mL and 2 mg/mL, respectively. Poly-TPD and PVK were spin-coated at 2000 rpm and annealed at 110°C for 20 minutes and 170°C for 30 minutes, respectively. The DiR layers were then spin-cast from toluene dispersion at 2000 rpm, and ligand-exchange and multilayer deposition was performed according the LbL procedure described above. As electron transport layer, ZnO NPs were spin-cast at 2500 rpm from ethanol solution. Finally, Al was deposited by thermal evaporation in a vacuum deposition chamber. The current–voltage–luminance characteristics were measured using a Keithley 2410 source-measure unit, and an Agilent 34410A multimeter coupled to a calibrated PDA 100A Si switchable gain detector from Thorlabs. The output of Si detector was converted into power (photon flux) using a 50 Ω load resistance and the responsivity of the detector. The EQE was calculated as the ratio of the photon flux and the driving current of the device. The EL spectra of the devices were obtained using an Ocean Optics HR4000+ spectrometer.

**Supporting Information Available**. (1) Structure of the ligands; (2) Optical data from nanocrystal solutions; (3) Details on the fitting of the PL decay traces; (4) Details on the layer-by-layer deposition process; (5) Cross section SEM, device statistics, and shelf life data of LEDs;

SYNOPSIS TOC

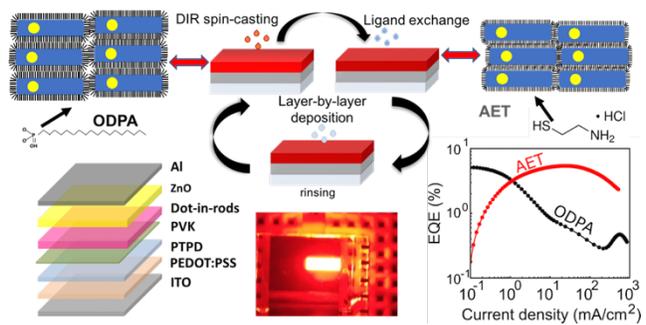